\documentclass{ws-mpla}

\begin{document}

\markboth{R. N. Panda, Mahesh K. Sharma and S. K. Patra}
{Nuclear Structure and Reaction Studies}

\catchline{}{}{}{}{}

\title{Nuclear Structure and Reaction Properties of Ne, Mg and Si Isotopes with RMF Densities}

\author{R. N. Panda}
\address{Department of Physics, ITER, Siksha O Anusandhan University,
Bhubaneswar-751 030, India 
\footnote{R. N. Panda,
Email:rnpanda@iopb.res.in}\\
}
\author{Mahesh K. Sharma}
\address{School of Physics and Material Science, Thaper
University, Patila-147 004, India
\footnote{Mahesh K. Sharma
Email:maheshphy82@gmail.com}\\
}
\author{S. K. Patra}
\address{Institute of Physics, Sachivalaya Marg, Bhubaneswar-751005,
India
\footnote{S. K. Patra
Email:patra@iopb.res.in}\\
}

\maketitle

\pub{Received (Day Month Year)}{Revised (Day Month Year)}

\begin{abstract}
We have studied nuclear structure and reaction properties of Ne, Mg and Si
isotopes, using relativistic mean field densities, in the frame work of
Glauber model. Total reaction cross section $\sigma_{R}$ for Ne isotopes on
$^{12}C$ target have been calculated at incident energy 240 MeV. The results are
compared with the experimental data and with the recent theoretical study
 [W. Horiuchi et al., Phys. Rev. C, $\bf 86$, 024614 (2012)]. Study of
$\sigma_{R}$ using deformed
densities have shown a good agreement with the data. We have also
predicted total reaction cross section $\sigma_{R}$ for Ne, Mg and Si isotopes
as projectiles and $^{12}C$ as target at different incident energies.
\end{abstract}

\ccode{21.10.Dr., 21.10.Ft, 21.10.Gv, 25.70.-z}

\section{Introduction}
The development of accelerator technique for Radioactive Ion Beams (RIBs)
help to study numerous experimental as well as theoretical measurements for
nuclei far from  $\beta$ - stability line. Experimental methods and theoretical
analysis have been widely used to collect information about the nuclear size,
 valence nucleon distribution and halo structure. The measurement of various
cross sections like reaction cross section $\sigma_R$, nucleon removal cross
section $\sigma_{-1n}$ and longitudinal momentum distribution $P_{||}$ are some of the established tools for exploring unstable nuclei. Island of inversion
(IOI) is one of the most important current subjects in nuclear physics. This was first applied by Warburton to a region of very neutron-rich nuclei from
$^{30}$Ne to $^{34}$Mg \cite{war90}. Discovery of the halo structure is another important progress of research on unstable nuclei. A halo structure of
$^{31}$Ne was reported by the experiment on the one neutron removal reaction
\cite{nakamura09}. Experimentally this is the heaviest halo nucleus. The
formation of halo in a nucleus near the drip-line is due to the very small
binding of the valence particles. The quadrupole deformation of the halo is
determined by the structure of the weakly bound valence orbital and it does not depend on the shape of the core \cite{misu97}. On the other hand existence of
two nucleon halo is most unlikely in a deformed nucleus \cite{nunes05}. It is
shown by Nunes \cite{nunes05} with a variety of 3-body NN tensor force which
goes beyond the usual pairing in Hartree-Fock-Bogoliubov (HFB) and the
coupling due to core deformation/polarization significantly reduce the formation of 3-body Borromean systems. In a recent work \cite{zhou10}, halo phenomena
in deformed nuclei are analysed within deformed Relativistic Hartree Bogoliubov (RHB) theory and their finding in weakly bound $^{44}$Mg  nucleus indicates a
decoupling of the halo orbitals from the deformed core agreeing with the
conclusion of Ref. \cite{misu97}.

To develop consistent nuclear reaction systematics along with the nuclear
structure, several theoretical models
have been a matter of wide interest. In this context, the relativistic mean
field (RMF) or the effective field theory motivated RMF (E-RMF) models
provide the internal structure  or sub-structure information
of the nuclei through density distributions \cite{bhuyan11}, which are used
as input while calculating the observables in conjunction with Glauber model
\cite{skpatra1,sharma06,skpatra09,rnpanda10,rnpanda11,mahesh12}. A systematic
study of various
nuclear reaction cross sections, such as the total nuclear reaction cross
sections, differential elastic cross sections etc. enables us to know the
nuclear structure of unstable nuclei in detail. This will also help in
studying the formation of neutron-rich nuclei that are surrounded by a
high pressure or temperature. In the present paper, our aim is to calculate
the bulk properties, such as  binding energy (BE), root mean square charge
radius $r_{ch}$, and quadrupole deformation parameter $\beta_2$ for
Ne, Mg and Si isotopes in the RMF and E-RMF formalisms. Then we analyze the
total nuclear reaction cross section $\sigma_{R}$ for the scattering of
$^{20-32}$Ne from a $^{12}$C target at 240 MeV/nucleon
by using the densities obtained from the RMF formalisms
\cite{skpatra1,sharma06,skpatra09,rnpanda11} in the frame work of Glauber
model. We have also predicted total reaction cross sections for Ne, Mg and Si
cases at different incident energies.

The paper is designed as follows: The RMF/E-RMF formalisms and the reaction
mechanism in the framework of Glauber model are explained briefly in Section II. The results obtained from our calculations are discussed in Section III. In
this Section we intend to study the applicability of Glauber model in the
context of both stable and drip-line nuclei. Finally, a brief summary and
concluding remarks are given in the last Section IV.

\section{Theoretical Framework}
\label{sec:1}
The successful applications of RMF both in finite and infinite nuclear systems
make more popular of the formalism in the present decades.
The RMF model has been extended to the Relativistic Hartree-Bogoliubov (RHB)
and density functional approach to study the static and dynamic aspects of exotic nuclear structure \cite{vretenar05,meng06}. The use of RMF formalism for finite nuclei as well as infinite nuclear matter are well documented and details can
be found in \cite{patra91,serot86,pring96,lala97,sharma93} and
\cite{tang97,patra01,skpatra01,serot97} respectively.
The working expressions for the density profile and other related quantities
are available in
\cite{skpatra1,sharma06,skpatra09,rnpanda11,patra91,serot86,pring96,tang97,patra01,skpatra01,serot97}.
The details to calculate $\sigma_R$ using Glauber approach has been given by
R. J. Glauber \cite{gla59}. This model is based on the independent,
individual nucleon-nucleon ($NN$) collisions along the eikonal
\cite{abu03}. It has been used extensively to explain the observed total
nuclear reaction cross-sections for various systems at high energies. The
standard Glauber form for the total reaction cross-sections at high energies is expressed as \cite{gla59,kar75}:
\begin{equation}
\sigma _{R}=2\pi\int\limits_{0}^{\infty }b[1-T(b)]db \;,
\end{equation}
where $T(b)$ is the transparency function with impact parameter $b$.
The function $T(b)$ is calculated in the overlap region between the projectile
and the target assuming that the interaction is formed from a single $NN$
collision. It is given by

\begin{equation}
T(b)=\exp \left[ -\sum\limits_{i,j}\overline{\sigma }_{ij}\int d%
\vec{s}\overline{\rho }_{ti}\left( s\right) \overline{\rho }%
_{pj}\left( \left| \vec{b}-\vec{s}\right| \right)
\right] \;.
\end{equation}
The summation indices $i$ and $j$ run over proton and neutron and subscript
$p$ and $t$ refers to projectile and target, respectively. The experimental
nucleon-nucleon reaction  cross-section $\overline{\sigma }_{ij}$ varies with
energy. The $z$-integrated densities $\overline{\rho }(\omega )$ are defined as
\begin{equation}
\overline{\rho }(\omega )=\int\limits_{-\infty }^{\infty }\rho \left( \sqrt{%
\omega ^{2}+z^{2}}\right) dz \;,
\end{equation}
with $\omega ^{2}=x^{2}+y^{2}$.  The argument of $T(b)$ in Eq. (2) is
$\left| \vec{b}-\vec{s}\right|$, which stands for the impact parameter between
the $i^{th}$ and $j^{th}$ nucleons.

The original Glauber model was designed for high energy approximation. However, it was found to work reasonably well for both the nucleus-nucleus reaction and
the differential elastic scattering cross-sections over a broad energy range
\cite{chau83,mbuen84}. To include the low energy effects of $NN$ interaction,
the  Glauber model is modified to take care of the finite range effects in the
profile function and Coulomb modified trajectories \cite{abu03,pshukla03}.
The modified
$T(b)$ is given by \cite{abu03,bhagwat},
\begin{small}
\begin{equation}
T(b)=\exp
\left[-\int_{p}\int_{t}\sum
\limits_{i,j}\left[\Gamma _{ij} \left( \vec{b} - \vec{s} + \vec{t}
\right) \right] \overline{\rho}_{pi} \left( \vec{t} \right)
\overline{\rho }_{tj} \left( \vec{s}\right) d\vec{s}d\vec{t} \right].
\label{eq:2}
\end{equation}
\end{small}
\noindent The profile function $\Gamma_{ij}(b_{eff})$ is defined as
\cite{skpatra1,sharma06}
\begin{equation}
\Gamma _{ij}(b_{eff})=\frac{1-i\alpha _{NN} }{2\pi \beta _{NN}^{2}}\sigma
_{ij}\exp \left( -\frac{b_{eff}^{2}}{2\beta _{NN}^{2}}\right)\;,
\label{eq:3}
\end{equation}
with $b_{eff}=\left| \vec{b}-\vec{s}+\vec{t%
}\right| $,  $\vec{b}$ is the impact parameter, $\vec{s}$ and $\vec{t}$ are
the dummy variables for integration over
the $z$-integrated target and projectile densities. The parameters $\sigma _{NN}$, $\alpha _{NN}$, and $\beta _{NN}$ are usually case-dependent (proton-proton, neutron-neutron or proton-neutron), but we have used the appropriate average
values from
Refs. \cite{kar75,charagi90,charagi92,charagi93,charagi97}.
It is worth mentioning that the result in Glauber model is sensitive to the
in-medium NN cross-section with proper treatment of the input densities
\cite{hussein91} and also depends on the accuracy of the profile function.

At intermediate energies, medium effects can be taken into account on
nucleon-nucleon cross-sections. In NN scattering the basic input is the NN
elastic t-matrix. This t-matrix is modified to take into account nuclear medium effects in both projectile and target. A. Bertulani et.al \cite{bertulani10}
have shown that the nucleon knockout reactions involving halo nuclei are more
sensitive to medium modifications compared to normal nuclei. The deformed or
spherical nuclear densities obtained from the RMF model are fitted to a sum of
two Gaussian functions with suitable co-efficients $c_i$ and ranges $a_i$ chosen for the respective nuclei which is expressed as
\begin{equation}
\rho (r)=\sum\limits_{i=1}^{2}c_{i}exp[-a_{i}r^{2}].
\end{equation}
The deformed intrinsic RMF densities are converted to its spherical equivalent
using this equation which is consistent with the Glauber theory applied in the
laboratory frame \cite{abu03}. Then, the Glauber model is used to calculate the total reaction cross-section for both the stable and unstable nuclei considered in the present study. In Refs. \cite{skpatra09,rnpanda10,abu03,pshukla03,bhagwat} it is shown that the Glauber model can be used for relatively low energy even
at 25, 30 and 85 MeV/nucleons. Although it is a better prescription to take deformation into account directly through the transparency function [Eqn. (4)], but to our knowledge no such scheme is available in this model. In our earlier
calculations \cite{skpatra1,sharma06,skpatra09,rnpanda11} we have used the
present approach to take deformation into account where the results are quite
encouraging and show clear deformation effect. It is to be noted that similar
methodology is also adopted by some other authors \cite{bhagwat}. Also, it is
important to note that the densities for halo-nuclei have long tails which
generally are not reproduced quantitatively by harmonic oscillator expansion in a mean field formalism.
\begin{figure}[th]
\vspace*{0.7cm}
\centerline{\psfig{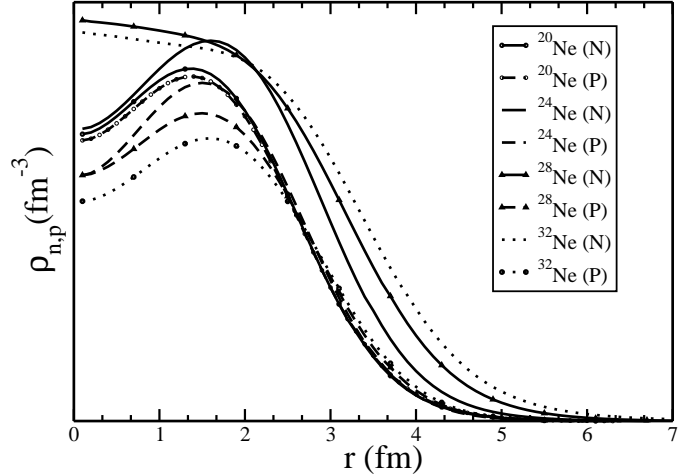}}
\vspace*{6pt}
\caption{The spherical proton ($\rho_p$) and neutron ($\rho_n$) density
obtained from RMF (NL3*) parameter set for various isotopes of Ne.
\protect\label{fig1}}
\end{figure}

\begin{figure}[th]
\centerline{\psfig{file=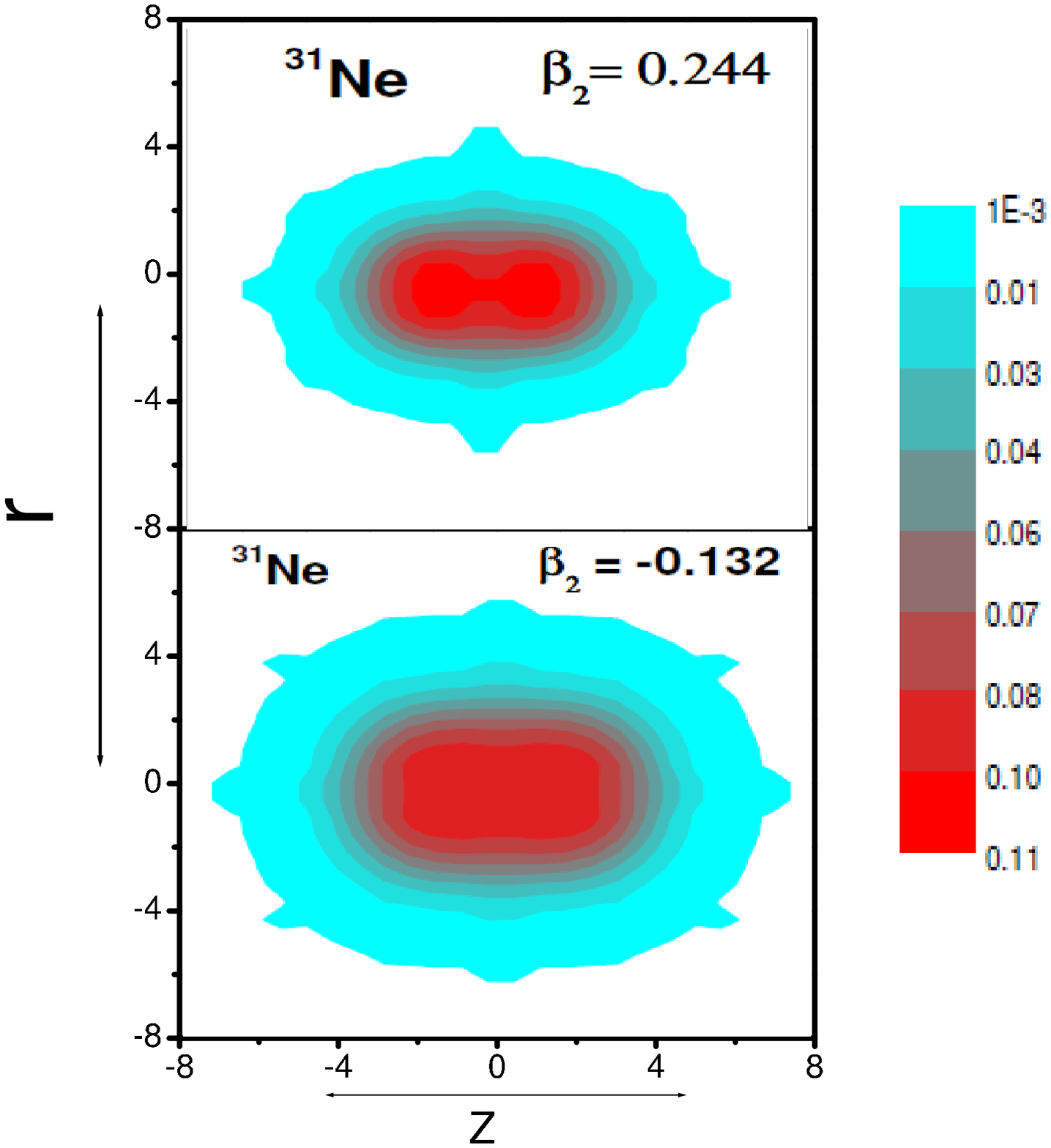,width=11cm}}
\vspace*{8pt}
\caption{The axially deformed density distribution for $^{31}$Ne with RMF (NL3*) parameter set. The width and height of the boxes are 8 fm each with the
uniform contour spacing of 0.01 $fm^{-3}$.
\protect\label{fig1}}
\end{figure}
\section{Calculations and Results}

We obtain the field equations for nucleons and mesons from the RMF and E-RMF
Lagrangian. For the deformed case (RMF only), these equations are solved by
expanding the upper and lower components of the Dirac spinners and the boson
fields in an axially deformed harmonic oscillator basis. The set of coupled
equations are solved numerically by a self-consistent iteration method taking
different inputs of the initial deformation $\beta_0$
\cite{patra91,serot86,pring96,gam90}. For spherical densities, we follow the
numerical procedure of Refs. \cite{patra01,skpatra01} for both RMF and E-RMF
models. The constant gap BCS pairing is used to add the pairing effects for
open shell nuclei. In the present calculations, we have dealt reaction studies
for nuclei Ne, Mg and Si with C target. All these nuclei are in the lower
region of the mass table, where the contribution of pairing effect is minimal
even for open shell nuclei \cite{bohr75}. We also understand that pairing plays
a crucial role for open shell nuclei for relatively heavier mass region. If
one use the conventional pairing gaps similar to $\triangle$ = 11.2/$\sqrt{A}$
MeV, then BCS treatment of pairing is not reliable. However, using small pairing gap near the dripline \cite{skpatra01,madland81}, this error can be minimised. We have used this scheme in our earlier calculations \cite{skp09} and
able to reproduce the results with data till the dripline whenever available.
The centre-of-mass motion (c.m.) energy correction is estimated by the
usual harmonic oscillator formula $E_{c.m.}=\frac{3}{4}(41A^{-1/3})$.

Since the main input in the Glauber model estimation is the RMF or E-RMF
densities, it is important to have information about these quantities. We have
plotted the spherical $\rho_p$ and $\rho_n$ for both proton and neutron
distributions of Ne isotopes in Figure 1 using RMF (NL3*) parameter set
\cite{lala97,sharma93}. As expected, we find the values of $\rho_n$ and $\rho_p$ are
almost similar for $^{20}$Ne
 which can be seen from Figure 1. Extension of $\rho_n$ is much more than
$\rho_p$ for rest of the nuclei. It is maximum for $^{32}$Ne in Neon isotopic
chain, because of high neutron to proton ratio for these cases.
The axially deformed density for the halo case $^{31}$Ne is shown
in Figure 2. The z-axis is chosen as the symmetry axis, the density is
evaluated in the $z\rho$ plane, where $\sqrt{x^2+y^2} = \rho$. In Ref. \cite{aru05}, it is noticed that $^{31}$Ne possesses a $3\alpha-$
cluster with a tetrahedral configuration. The structure of this neutron-rich
$^{31}$Ne has an prolate ground state deformation. The density plot shows that
the central part of the nucleus is a compact core, which is surrounded by a
thin layer of nucleons.

\begin{table}[h]
\tbl{Calculated results for binding energy (BE), root mean square charge
radius $r_{ch}$, and quadrupole deformation parameter $\beta_2$ for the neutron rich $^{18-32}$Ne isotopes using RMF and E-RMF densities obtained from NL3*
and G2 parameter sets respectively. The available experimental data
are given for the comparison. BE is in MeV and  $r_{ch}$ in fm.}
{\begin{tabular}{@{}cccccccccccc@{}}
\toprule
Nucleus &\multicolumn{3}{c|}{BE}&\multicolumn{3}{c|}{$r_c$}&\multicolumn{3}{c|}
{$\beta_2$}\\
\colrule
&RMF& E-RMF& Expt. &RMF&E-RMF& Expt.&RMF&Ref. \cite{sumi12}& Expt.\\

\colrule
$^{18}$Ne&131.8&135.3&132.1&2.963&3.055&2.972&0.238&&0.68(3) \\
$^{20}$Ne&156.7&156.6&160.6&2.972&2.986&3.00&0.537&0.479&0.70(20) \\
$^{22}$Ne&175.7&174.2&177.8&2.94&2.903&2.954&0.502&0.400&0.564(4) \\
$^{24}$Ne&189.1&190.2&191.8&2.88&2.879&2.903&$-$0.259&0.258&0.41(5) \\
$^{26}$Ne&200.0&202.7&201.5&2.926&2.886&2.927&0.277&0.221&0.39(3)  \\
$^{28}$Ne&208.3&211.7&206.9&2.965&2.925&2.963&0.225&0.526&0.36(3) \\
$^{30}$Ne&215.2&218.2&211.3&2.992&2.965&&0.046&0.400&0.49(17) \\
$^{31}$Ne&216.3&220.0&211.6&3.027&2.974&&0.244&0.422& \\
$^{32}$Ne&218.7&221.2&213.2&3.069&2.982&&0.369&0.335& \\
\botrule
\end{tabular}\label{ta1}}
\label{Table 2}
\end{table}

\begin{table}[h]
\tbl{Same as table 1 but for $^{22-34}$Mg isotopes.}
{\begin{tabular}{@{}cccccccccccc@{}}
\toprule
Nucleus &\multicolumn{3}{c|}{BE}&\multicolumn{3}{c|}{$r_c$}&\multicolumn{2}{c|}
{$\beta_2$}\\
\colrule
&RMF& E-RMF& Expt. &RMF&E-RMF& Expt.&RMF& Expt.\\

\colrule
$^{22}$Mg&166.42&165.63&168.58&3.092&3.142&&0.5128&0.65(12)  \\
$^{24}$Mg&194.31&189.44&198.26&3.043&3.037&3.057&0.4874&0.613(14)  \\
$^{26}$Mg&212.54&211.20&216.68&2.978&2.982&3.033&0.2728&0.484(6)  \\
$^{28}$Mg&228.76&228.45&231.63&3.048&3.011&&0.3447&0.484(20) \\
$^{30}$Mg&240.51&241.68&241.64&3.062&3.042&&0.2406&0.41(3)  \\
$^{32}$Mg&250.59&252.69&249.81&3.090&3.076&&0.1190&0.51(5)  \\
$^{34}$Mg&257.39&259.47&256.48&3.150&3.091&&0.3432&0.55(6)  \\
$^{36}$Mg&264.13&264.08&260.80&3.198&3.102&&0.4344&  \\
\botrule
\end{tabular}\label{ta1}}
\label{Table 2}
\end{table}

\begin{table}[h]
\tbl{Same as table 1 but for $^{26-36}$Si isotopes.}
{\begin{tabular}{@{}cccccccccccc@{}}
\toprule
Nucleus &\multicolumn{3}{c|}{BE}&\multicolumn{3}{c|}{$r_c$}&\multicolumn{2}{c|}
{$\beta_2$}\\
\colrule
&RMF& E-RMF& Expt. &RMF&E-RMF& Expt.&RMF& Expt.\\

\colrule
$^{26}$Si&200.86&202.84&206.04&3.118&3.136&&$-$0.2800&0.444(21)  \\
$^{28}$Si&232.13&230.54&236.54&3.122&3.065&3.122&$-$0.3308&0.412(4)  \\
$^{30}$Si&250.58&251.55&255.62&3.070&3.09&3.133&0.1481&0.330(22)  \\
$^{32}$Si&268.45&269.25&271.41&3.137&3.116&&$-$0.2007&0.26(4)  \\
$^{34}$Si&284.45&285.05&283.43&3.147&3.152&&0.0005&0.18(4)  \\
$^{36}$Si&291.57&295.59&292.03&3.186&3.166&&$-$0.1616&0.25(4)  \\
\botrule
\end{tabular}\label{ta1}}
\label{Table 2}
\end{table}

\begin{figure}[th]
\vspace*{0.7cm}
\centerline{\psfig{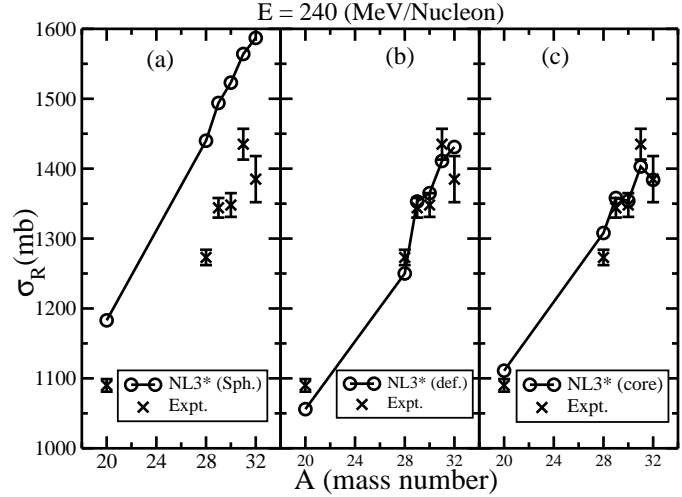}}
\vspace*{6pt}
\caption{Calculated reaction cross sections for scattering of Ne isotopes
on $^{12}$C target at 240 MeV/nucleon with experimental data.
\protect\label{fig1}}
\end{figure}

\begin{figure}[th]
\vspace*{0.5cm}
\centerline{\psfig{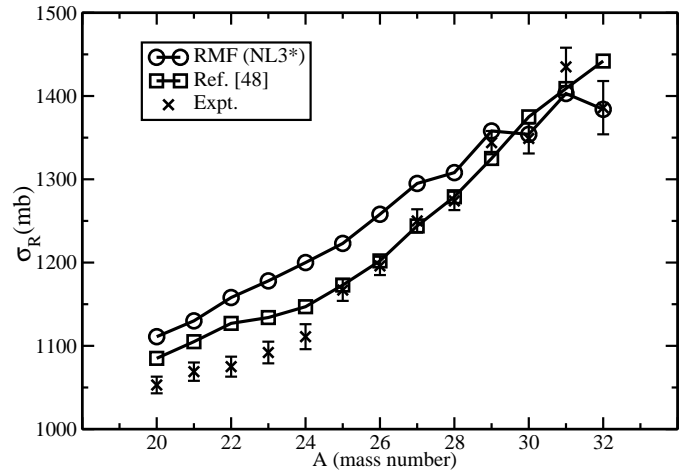}}
\vspace*{6pt}
\caption{Comparison of our results with W. Horiuchi et al., Phys. Rev. C,
$\bf 86$, 024614 (2012) and experimental data at 240 MeV/nucleon for scattering of $^{20-32}$Ne isotopes on $^{12}$C target.
\protect\label{fig1}}
\end{figure}

\begin{figure}[th]
\centerline{\psfig{file=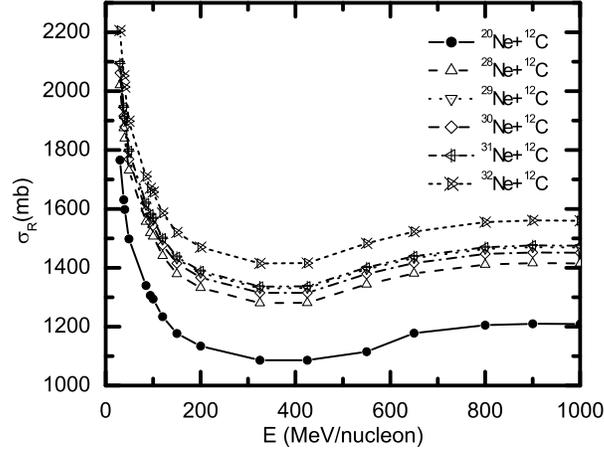,width=8.9cm}}
\vspace*{6pt}
\caption{Same as Figure 3 but for different incident energies.
\protect\label{fig2}}
\end{figure}

\begin{figure}[th]
\centerline{\psfig{file=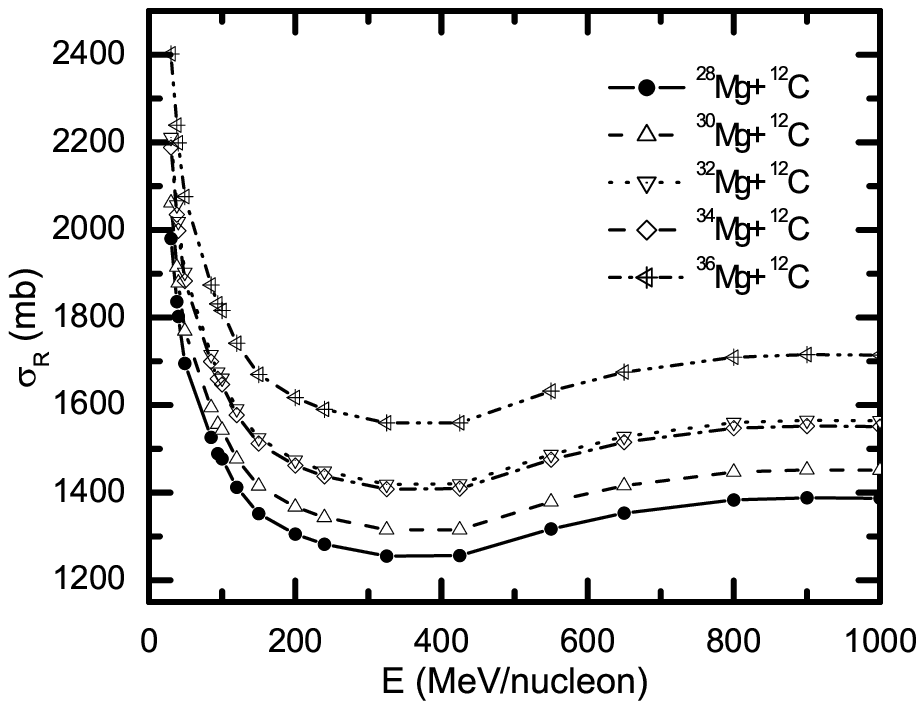,width=8.9cm}}
\vspace*{6pt}
\caption{Reaction cross sections for scattering of Mg isotopes
on $^{12}$C target at different incident energies.
\protect\label{fig3}}
\end{figure}

\begin{figure}[th]
\centerline{\psfig{file=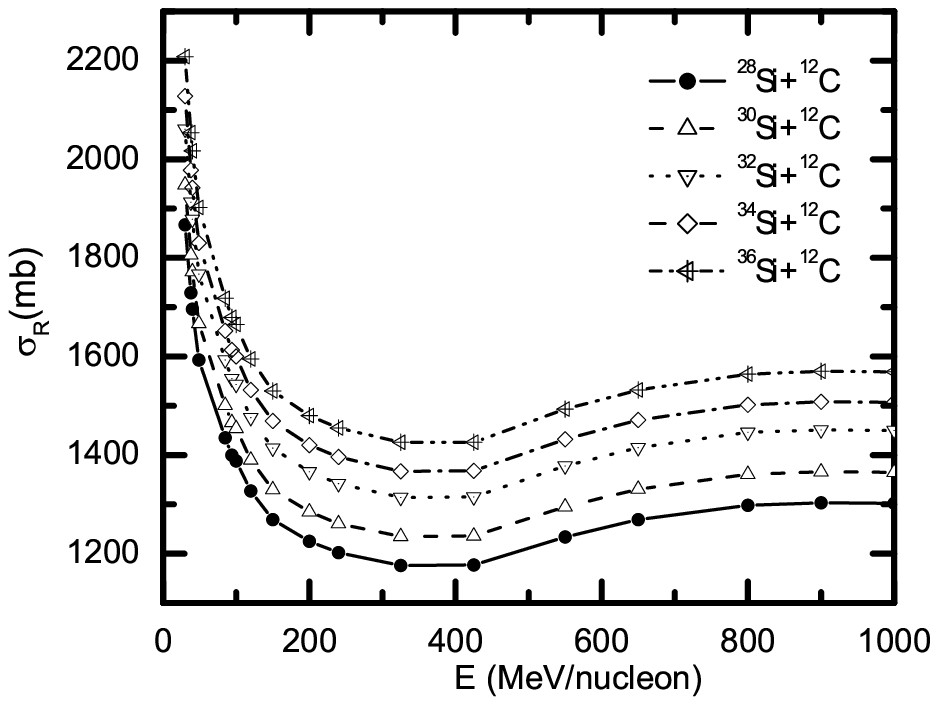,width=8.9cm}}
\vspace*{6pt}
\caption{Reaction cross sections for scattering of Mg isotopes
on $^{12}$C target at different incident energies.
\protect\label{fig3}}
\end{figure}

\begin{figure}[th]
\centerline{\psfig{file=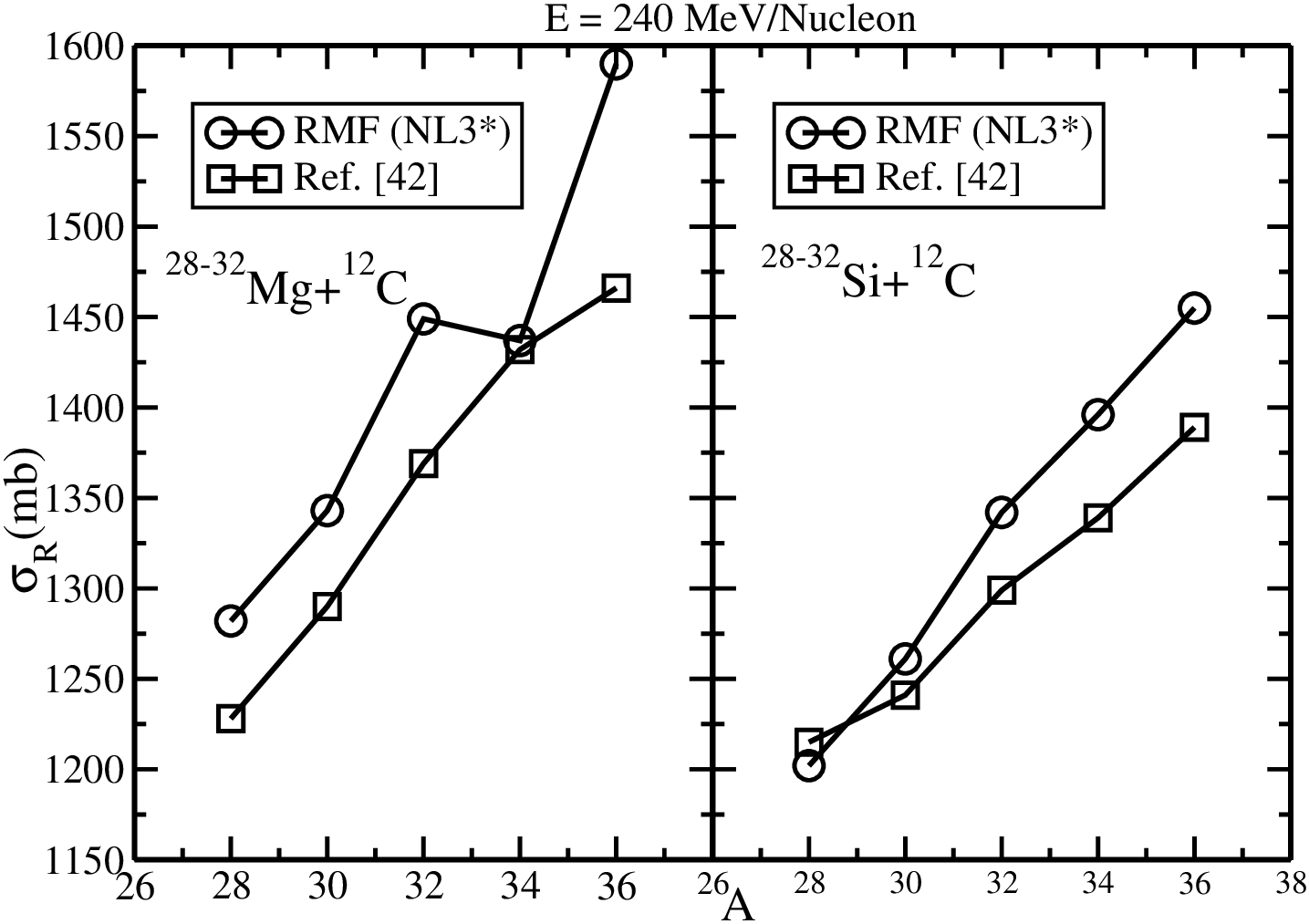,width=8.9cm}}
\vspace*{6pt}
\caption{Comparison of our results with W. Horiuchi et al., Phys. Rev. C,
$\bf 86$, 024614 (2012) and experimental data at 240 MeV/nucleon for scattering of $^{20-32}$Ne isotopes on $^{12}$C target.
\protect\label{fig3}}
\end{figure}

We calculate the bulk properties, such as binding energy (BE), root mean
square charge radius $r_{ch}$, and quadrupole deformation parameter $\beta_2$
for the neutron rich $^{18-32}$Ne, $^{24-34}$Mg and $^{26-36}$Si isotopes in
the RMF and E-RMF formalisms. The calculated nuclear structure results are
compared with T. Sumi
et al. \cite{sumi12} and available experimental data \cite{audi03,audi11,data} 
in
Tables 1, 2 and 3. It is
clear that our results agree remarkably well with the data. For example, the
RMF binding energy for $^{18}$Ne is 131.8 MeV and 135.3 MeV from E-RMF where as the experimental value is 132.1 MeV. Similarly, the $r_{ch}$ value for this
nucleus is 2.963, 3.055 and 2.972 fm for RMF, E-RMF and experiment respectively. A comparison with the study of T. Sumi et al. \cite{sumi12} for the deformation parameter $\beta_2$ is also given in Table 1. We found that the results are
quite close with each other.

In the present study of $\sigma_R$, we first use the spherical density obtained from RMF (NL3*)\cite{lala97,sharma93}. The results are presented in Figure 3(a) for
$^{20}Ne$ and $^{28-32}Ne$ isotopes with $^{12}C-$ target at 240 MeV/Nucleon
projectile energy. These results deviate considerably from the data
\cite{takechi10,takechi10a,takechi12} which are quoted
in this figure. For example, in case of $^{28}$Ne+$^{12}$C, the observed value
of $\sigma_{R}$ is $1273\pm11$ MeV as compared to the estimated results of
1440 MeV with NL3* parametrization.

In this context, it is interesting enough to see deformation effect on
$\sigma_{R}$. We repeat the calculations for $\sigma_{R}$ with the deformed
densities (RMF only) as input in the Glauber model \cite{skpatra1,sharma06}.
We obtained spherical equivalent of the axially deformed densities using
equations (3) and (6) following the prescription of
Refs. \cite{skpatra1,sharma06,skpatra09,rnpanda11,mahesh12}. The NL3* parameter set
\cite{sharma93} for this purpose is used and the results are presented in
Figures 3(b) and 3(c). The parameter set NL3* is reasonably a good set
for these neutron-rich nuclei. It shows that most of the $\sigma_{R}$ matches
quite well with the experimental data of
\cite{takechi10,takechi12} and still halo
case does not agree in Figure 3(b). In Figure 3(c), we have
taken core + one neutron case and found a remarkable agreement with the
data. In figure 4, we have made a comparison of our results with W. Horiuchi
et al. \cite{horiuchi12} as well as with the data for the scattering of
$^{20-32}Ne$ isotopes on $^{12}C-$ target at 240 MeV/Nucleon. It is found that
our results are little bit higher for relatively lower mass nuclei but there is better fitting for neutron rich side. Summarising the whole discussion on
reaction cross sections, in general, one can say
that the spherical density used from RMF (NL3*) fails to reproduce the data.
When we use the deformed densities to evaluate the total nuclear reaction cross section, the predicted $\sigma_{R}$ matches reasonably well with the
experimental measurement. In Figures 5, 6 and 7, we have presented the
$\sigma_{R}$ with various incident energies for Ne, Mg and Si isotopes as
projectiles and $^{12}C$ as target using the deformed NL3* densities in the
Glauber model calculation. A comparison of oue results with W. Horiuchi et al.
\cite{horiuchi12} for the scattering of Mg and Si isotopes on $^{12}C-$ target
at 240 MeV/Nucleon is also given in Figure 8. It is observed a similar trend as
reported in \cite{horiuchi12}.

\section{ Summary and Conclusion}
In summary, the binding energy, charge radius and quadrupole deformation
parameter for the neutron-rich $^{18-32}Ne$, $^{24-34}$Mg and $^{26-36}$Si
isotopes have been calculated using RMF (NL3*) and E-RMF (G2) formalisms.
Using the RMF densities, the reaction cross sections $\sigma_{R}$ are evaluated
in the Glauber model. The
$\sigma_{R}$ are in good agreement with the experiments and also with the
recent study by W. Horiuchi et al \cite{horiuchi12}, when we consider the
deformation effect in the densities. It is also concluded in the present paper
that deformation for total nuclear reaction cross section is important
for stable nuclei as projectiles which reproduces the experimental data
reasonably well. In $^{31}Ne$ halo case, the deformed core + one neutron shows
better agreement with the data.

\section{Acknowledgement}
We are thankful to M. Bhuyan for his support during the work.

\end{document}